\documentclass[12pt,letterpaper]{JHEP3}
\pdfoutput=1
\usepackage{epsfig}
\usepackage{subcaption}
\usepackage{graphicx}
\usepackage[english]{babel}

\usepackage{cite}

\graphicspath{{Figures/}}

\title{The missing top of AdS resonance structure}

\author{I-Sheng Yang\\
ITFA and GRAPPA, Universiteit van Amsterdam, \\
Science Park 904, 1090 GL Amsterdam, Netherlands
}

\abstract{
We study a massless scalar field in $AdS_{d+1}$ with a nonlinear coupling $\phi^N$ and not limited to spherical symmetry. The free-field-eigenstate spectrum is strongly resonant, and it is commonly believed that the nonlinear coupling leads to energy transfer between eigenstates. We prove that when $Nd$ is even, the most efficient resonant channels to transfer energy are always absent. In particular, for $N=3$ this means no energy transfer at all. For $N=4$, this effectively kills half of the channels, leading to the same set of extra conservation laws recently derived for gravitational interactions within spherical symmetry.
}

\begin{document}

\section{Introduction and summary}

Anti-de Sitter space has many intriguing properties. One of those is a strongly resonant spectrum. For any massless field, for example gravitational waves, the eigenstates all have integer frequencies. When there is a nonlinear (self-)coupling, for example through gravitational back-reactions, such resonance allows energy to be transferred between different eigenstates. Thus, no matter how small the coupling is, after a correspondingly long time, the initial energy can end up being anywhere, leading to rich and unpredictable dynamics.

Some authors have argued that these many channels to transfer energy generically leads to energy cascade: starting from a few low frequency modes, energy continuously spreads out into higher ones. This naturally accumulates to a significant effect at the time scale set by the inverse coupling strength. This ``resonant-cascade'' theory has been widely quoted as the explanation for a nonlinear gravitational instability: black hole formation often observed in numerical simulations at this time scale \cite{BizRos11,DiaHor11,MalRos13a,BizJal13,HorSan14}. 

However, as emphasized in \cite{CraEvn14,DFLY14}, the existence of these channels is only a necessary condition for energy cascade, and energy cascade is again only a necessary condition for black hole formation\footnote{The AdS (in)stability problem is indeed one important motivation of this work, but many related articles are not directly relevant for this paper. For a more complete list of reference about the (in)stability problem, please see \cite{DFLY14}.}. Since it is doubly insufficient, the resonant cascade theory cannot be a good dynamical explanation of why black hole forms in those cases. In other words, the resonant-cascade theory only derives a lowerbound of the time scale in which a black hole can form.  It does not really explain how a black hole forms, and why in some cases it happens so fast, almost saturating this lowerbound. Such dynamical understanding likely requires one to go beyond any theory in the eigenstate spectrum and directly study the position space behavior\cite{DFLY14}.

On the other hand, since it is a necessary condition, analyzing the energy flow between eigenstates can provide excellent arguments against black hole formation. This has been a fruitful line of thoughts. For example, it was argued that a small modification of the AdS space makes the spectrum only asymptotically resonant, which will be not enough for energy cascade\cite{DiaHor12}. Later several numerical simulations indeed confirmed that\cite{MalRos14}. It was also observed that half of the na\"ively expected channels to transfer energy are actually absent, which leads to extra conserved quantities. This may explain the stable, quasi-periodic solutions which are also often observed during numerical simulations\cite{BalBuc14,CraEvn14,CraEvn14a,BucGre14}.

In this paper, we follow this fruitful line of thoughts. The missing channels and the consequent conservation laws were established within spherical symmetry, but an explicit evaluation of several low energy eigenstates demonstrated the same behavior even without spherical symmetry\cite{HorSan14}\footnote{Although the former is done for a scalar field but the later is for pure gravity, the general structure of the eigenstates are the same.}. Thus, a natural next step is to establish the missing channels beyond spherical symmetry. We study a massless scalar field with a general nonlinear coupling, $\phi^N$ with $N\geq3$, and in $AdS_{d+1}$ with $d\geq2$. We will show that the missing channels are generic for AdS eigenstates, and those missing are exactly the most efficient ones for transferring energy into higher eigenstates.

More technically, the energy spectrum of a massless scalar field is given by
\begin{equation}
w_{nl} = 2n + l + d~,
\end{equation} 
where the radial wavenumber $n$ and the total angular momentum $l$ are both non-negative integers. The resonant condition is
\begin{equation}
w_{n_1l_1} = \sum_{i=2}^N \pm w_{n_il_i}~,
\label{eq-res}
\end{equation}
with $(N-1)$ arbitrary choices of $\pm$ while maintaining the positivity of frequencies. Basically, the $(N-1)$ eigenstates on the r.h.s. can conspire to transfer energy to the eigenstate on the l.h.s. if this condition is met\footnote{Of course, certain angular momentum summation rules must also be satisfied between these eigenstates, but there is a large degeneracy involved so it can always be done, and we will ignore such complication in this paper.}. If we choose all $+$ signs in this resonant condition, we get 
\begin{equation}
w_{n_1l_1} = \sum_{i=2}^N  w_{n_il_i}~.
\label{eq-topintro}
\end{equation}
These channels represent the most efficient ways to transfer energy into high energy states if we start with only low energy ones. We call it the ``top'' of the AdS resonance structure.

In Sec.\ref{sec-proof}, we explicitly prove that these channels are absent whenever $Nd$ is even. In particular, when $N=3$, all resonant channels take this form, so this implies no energy transfer at all between eigenstates. The $\phi^3$ coupling only introduces subleading corrections to the free eigenstates, and any solution can still be expressed as a superposition of these approximate eigenstates. It is rather intriguing that such behavior is limited to even spatial dimensions.

In Sec.\ref{sec-N=4} we demonstrate that when $N=4$, this ``missing channels'' property is a direct generalization of what previously proven within spherical symmetry for weak gravitational self-interaction\cite{CraEvn14}. Based on the frequently observed similarity between $\phi^4$ theory and weak gravity, also the explicit evaluation of several coupling terms of the gravity theory without spherical symmetry\cite{HorSan14}, it is very likely that general gravitational interactions also have these channels missing. As a first step toward such generalization, we provide a slightly simpler proof of the conservation laws recently pointed out in\cite{BasKri14,CraEvn14a,BucGre14}, such that their non-spherically symmetric versions are obviously also valid. 

Finally, in Sec.\ref{sec-dis}, we emphasize two particular properties which suggest that there should be a more elegant group theory method to prove the ``missing top''. For gravitational interactions without spherical symmetry, it is probably wiser to formulate such method instead of trying to explicitly evaluate the couplings between eigenstates.


\section{The missing top}
\label{sec-proof}

\subsection{The general proof}

Consider the global $AdS_{d+1}$ spacetime,
\begin{equation}
ds^2 = \frac{1}{\cos^2x}
\left(-dt^2+dx^2+\sin^2x~d\Omega_{d-1}^2\right)~,
\end{equation}
and a massless scalar field with a nonlinear self-coupling,
\begin{equation}
S = \int 
\frac{g_{\mu\nu}}{2}
\partial^\mu\phi\partial^\nu\phi 
+ \frac{\phi^N}{N}~
\sqrt{g} dV_{d+1}~.
\end{equation}
This leads to the equation of motion
\begin{equation}
-\ddot\phi + \frac{1}{\cos^2x}\hat{L}_d\phi = \frac{\phi^{N-1}}{\cos^2x}~,
\label{eq-eom}
\end{equation}
where $\hat{L}_d$ is the Laplacian operator for the spatial metric. Ignoring the coupling term on the r.h.s., the free field solutions can be decomposed into separable eigenstates\cite{HamKab06,HamKab06a}.
\begin{eqnarray}
\phi_0(x,t,\Omega_{d-1}) &=& \sum_{n,l,\vec{m}} 
\left(A_{nl\vec{m}}e^{-iw_{nl}t}+\bar{A}_{nl\vec{m}}e^{iw_{nl}t}\right)
~e_{nl\vec{m}}(x,\Omega_{d-1})~, \\
w_{nl} &=& 2n + l + d~, \\
e_{nl\vec{m}}(x,\Omega_{d-1}) &=&
\cos^dx~\sin^lx~Y_{l\vec{m}}(\Omega_{d-1})~P_n^{(d/2+l-1,d/2)}(\cos2x)~.
\end{eqnarray}
$P_n^{\alpha,\beta}$ is the Jacobi polynomial, $l$ is the magnitude of the total angular momentum, $\vec{m}$ describes its components, and $Y_{l\vec{m}}$ is the generalized spherical harmonics. These eigenstates form an orthogonal basis,
\begin{eqnarray}
\int e_{n_il_i\vec{m}_i}e_{n_jl_j\vec{m}_j}\tan^{d-1}x
~dxd\Omega_{d-1} 
\propto \delta_{n_in_j}\delta_{l_il_j}
\delta_{\vec{m}_i\vec{m}_j}~.
\end{eqnarray}

In general, the nonlinear coupling allows energy to be transferred between these eigenstates. One can model that by a perturbative expansion also in the eigenstate basis,
\begin{eqnarray}
\phi &=& \phi_0 + \phi_1 + ..., \ \ \ \ \ 
\phi_1 = \sum_{n,l,\vec{m}} c_{nl\vec{m}}(t)
e_{nl\vec{m}}~, \label{eq-naive} \\
\ddot{c}_{n_1l_1\vec{m}_1}&+&
w_{n_1l_1}^2c_{n_1l_1\vec{m}_1}
= \sum_{n_j,l_j}^{2\leq j\leq N} S_{\{nl\vec{m}\}} \prod_{j=2}^N
\left(A_{n_jl_j\vec{m}_j}e^{-iw_{n_jl_j}t}+\bar{A}_{n_jl_j\vec{m}_j}e^{iw_{n_jl_j}t}\right)~, \\
S_{\{nl\vec{m}\}} &\equiv&
S_{n_1l_1\vec{m}_1n_2l_2\vec{m}_2...n_Nl_N\vec{m}_N}
= \int 
\prod_{i=1}^N e_{n_il_i\vec{m}_i}~
\tan^{d-1}x \frac{dxd\Omega_{d-1}}{\cos^2x}~.
\label{eq-channel}
\end{eqnarray}
Through the nonzero coupling coefficients $S_{\{nl\vec{m}\}}$, combinations of $(N-1)$ zeroth order modes source the first order correction in one mode.

Of course, many of these coefficients will be zero from the integral of eigenstates. We will worry about those later. First we should note that we do not care too much about some of them even if they are not zero. If the resonant condition, Eq.~(\ref{eq-res}), is not satisfied, then these combinations of $(N-1)$ modes drive the $c_{n_1l_1\vec{m}_1}$ harmonic oscillator not at resonance. If those are the only nonzero coefficients, then for a small zeroth order magnitude, we can have $\phi_1\sim\phi_0^{N-1}\ll\phi_0$, and the nonlinear coupling simply leads to small corrections of the free eigenstates. Most of the energy stays within the original eigenstates.

When Eq.~(\ref{eq-res}) is satisfied, those combinations drive the oscillator at resonance, thus leading to a secular growth of $c_{n_1l_1\vec{m}_1}$. In this case, $\phi_1$ will soon become comparable to $\phi_0$, and the na\"ive perturbation theory breaks down. In other words, a significant amount of energy is transferred from the original eigenstates into others. Now, the interesting question is how many of these coefficients satisfying the resonant condition are actually nonzero? Here we will prove that when the sign choices in the resonant condition are all $+$, namely in the form of Eq.~(\ref{eq-topintro}), then those coefficients are all zero.

We first rewrite Eq.~(\ref{eq-topintro}) more explicitly as
\begin{equation}
2n_1+l_1 = (N-2)d + \sum_{i=2}^{N} (2n_i+l_i)~,
\label{eq-top}
\end{equation}
and also the explicit integral of the coupling coefficients,
\begin{eqnarray}
S_{\{nl\vec{m}\}} &=&
\int \left(\prod_{i=1}^N Y_{l_i\vec{m}_i}\right)
~d\Omega_{d-1}
\\ \nonumber
& & \int_0^{\pi/2} 
\left(\prod_{i=1}^N \sin^{l_i}x~P_{n_i}^{d/2+l_i-1,d/2}\right)\cos^{Nd-2}x~\tan^{D-1}x~dx~.
\end{eqnarray}
The $\Omega_{d-1}$ integral leads to $(d-2)$ conditions for matching the angular momentum components and the generalized triangular inequality for total angular momentum, $l_i \leq \sum_{j\neq i}l_j$. Neither will be very important for our main purpose. Our proof only requires the $x$ integral.

Changing the variable to $y=\cos2x$, the $x$ integral becomes
\begin{eqnarray}
\int_{-1}^1 \left(\prod_{i=1}^N P_{n_i}^{d/2+l_i-1,d/2}(y)\right) (1-y)^{L/2+d/2-1}(1+y)^{(N-1)d/2-1}dy~,
\end{eqnarray}
where $L=\sum_i l_i$. We then use the definition of the Jacobi polynomial,
\begin{equation}
P_n^{(\alpha,\beta)}(y)\propto
(1-y)^{-\alpha}(1+y)^{-\beta}
\frac{d^n}{dy^n}[(1-y)^{\alpha+n}(1+y)^{\beta+n}]~,
\end{equation}
to write down $P_{n_1}$ explicitly in the integral,
\newpage
\begin{eqnarray}
\int_{-1}^1 & & 
\left(\prod_{i=2}^N P_{n_i}^{d/2+l_i-1,d/2}(y)\right) (1-y)^{L/2-l_1}(1+y)^{(N-2)d/2-1} \nonumber \\
& & \frac{d^{n_1}}{dy^{n_1}}[(1-y)^{d/2+l_1+n_1-1}
(1+y)^{d/2+n_1}]~dy~.
\end{eqnarray}

We note that when Eq.~(\ref{eq-top}) is satisfied, $L$ and $Nd$ must be together even or odd. Thus when either $N$ or $d$ is even, the first line in the above integrand is a polynomial of $y$. Since integration by part produces no boundary terms, the condition for the integral to not vanish is
\begin{equation}
n_1 \leq \left(\sum_{i=2}^N n_i + \frac{l_i}{2}\right)
-\frac{l_1}{2}+\frac{(N-2)d}{2}-1~.
\label{eq-seln}
\end{equation}
Since this contradicts Eq.~(\ref{eq-top}), we have proven that $S_{\{nl\vec{m}\}}=0$ for the resonant channels satisfying Eq.~(\ref{eq-topintro}).

\subsection{$N=3$ and other odd numbers}
\label{sec-N=3}

For $N=3$, $w_{n_1l_1} = w_{n_1l_2} + w_{n_3l_3}$ is the only form a resonant condition can take. By proving that such coupling coefficients all vanish, we showed that no energy transfer actually occur despite a strongly resonant spectrum. The original eigenstates only receive small corrections through the coupling, and nothing dramatic will happen given any initial condition. Note that our proof only works when $d$ is even. When $d$ is odd, we explicitly evaluated some coefficients, and they are indeed nonzero.

For larger odd number $N$ and in even $d$ dimensions, these ``top'' resonant channels remain missing, but lower resonant channels, those with more ``$-$'' signs in Eq.~(\ref{eq-res}), do exist. We also explicitly evaluated some of those to confirm that. The radial integral does not seem to give rise to other constraints, which agrees with our physical intuitions. Other constraints will only come from the angular $\Omega_{d-1}$ integral. This ``missing top'' behavior is again only true in even $d$. We evaluated a few coefficients in odd $d$ and saw that these top channels do exist.

On the other hand, when $d$ is odd, something interesting already happens when limited to spherical symmetry. All the frequencies, with $l=0$, are odd numbers. When $N$ is odd, the resonant condition, Eq.~(\ref{eq-res}), involves an odd number of frequencies, thus cannot be satisfied independent of the those ``$\pm$'' sign choices. As a result, when both $N$ and $d$ are odd, if we start with spherically symmetric initial data, there is again no energy transfer at all. These interesting behaviors related to the parity of spatial dimensions might provide further insight to express and prove the ``missing top'' and other constraints in a more elegant formalism\footnote{We thank Luis Lehner for pointing out the possible connection to the Huygen-Fresnel principle, although we have not been able to make use of that further.}.

\section{$N=4$ and conserved quantities}
\label{sec-N=4}

\subsection{Extra conserved quantities}

For $N=4$, the resonant condition has two qualitatively different forms, either ``3-1'' or ``pairwise''.
\begin{eqnarray}
{\rm either} & & w_{n_1l_1} 
= w_{n2l_2} + w_{n_3l_3} + w_{n_4l_4}~, \\
{\rm or} & & w_{n_1l_1} + w_{n_2l_2} 
= w_{n_3l_3} + w_{n_4l_4}~.
\end{eqnarray}
Our result shows that the 3-1 channels do not exist, only the pairwise ones do. The same conclusion was reached in\cite{CraEvn14} when they studied gravitational self-interactions within spherical symmetry. More recently, combining the absence of these channels and the symmetry properties of the coupling coefficients, extra conserved quantities were found\cite{CraEvn14a,BucGre14}. Here we will revisit the proof of those conservation laws and show that they are also valid in the $\phi^4$ theory even without spherical symmetry.

We first follow the two-time formalism introduced in\cite{BalBuc14}. Instead of the na\"ive perturbative expansion in Eq.~(\ref{eq-naive}), we assume that the amplitudes of the eigenstates are also time dependent, but evolve much more slowly.
\begin{eqnarray}
\phi(x,t,\Omega_{d-1}) &=& \sum_{n,l,\vec{m}}
\left(A_{nl\vec{m}}(t)e^{-iw_{nl}t}+A_{nl\vec{m}}(t)e^{iw_{nl}t}\right)
e_{nl\vec{m}}(x,\Omega_{d-1})~, \\
|\dot{A}_{nl\vec{m}}|^2 &\ll& 
w_{nl}^2|A_{nl\vec{m}}|^2~.
\end{eqnarray}
This assumption will not easily break down as the na\"ive perturbation theory. The leading order effect of the coupling only requires us to solve
\begin{eqnarray}
-2iw_{n_1l_1} \frac{dA_{n_1l_1\vec{m}_1}}{dt}&=& 
\sum_{n_2l_2\vec{m}_2}\sum_{n_3l_3\vec{m}_3}
\sum_{n_4l_4\vec{m}_4} S_{\{1234\}} 
\bar{A}_{n_2l_2\vec{m}_2}
A_{n_3l_3\vec{m}_3}A_{n_4l_4\vec{m}_4}~,
\\
S_{\{1234\}}&\equiv&
S_{\{n_1l_1\vec{m}_1\}\{n_2l_2\vec{m}_2\}
\{n_3l_3\vec{m}_3\}\{n_4l_4\vec{m}_4\}}~.
\end{eqnarray}
The solutions $A_{nl\vec{m}}(t)$ then models how energy is being slowly transferred between eigenstates. As the subscript becomes too long, we will resort to the above abbreviation.

While limited to $l=\vec{m}=0$, in\cite{CraEvn14a,BucGre14} it was shown that the symmetry properties of $S_{1234}=S_{2134}=S_{3412}$ leads to a conserved ``particle number''
\begin{equation}
\frac{d}{dt} \sum_{n} w_{n} 
|A_{n}|^2 
= \frac{-1}{2i}\sum_{\{1234\}}S_{\{1234\}}
\left(\bar{A}_{n_1}\bar{A}_{n_2}A_{n_3}A_{n_4}-A_{n_1}A_{n_2}\bar{A}_{n_3}\bar{A}_{n_4}\right)=0~.
\end{equation}
This is simply because under the exchange of $\{12\}\leftrightarrow\{34\}$, $S_{\{1234\}}$ is symmetric but the next factor is antisymmetric.
These symmetries are still there including eigenstates of nonzero angular momenta, so it is straightforward to verify that
\begin{equation}
\frac{d}{dt} \sum_{nl\vec{m}} w_{nl\vec{m}} 
|A_{nl\vec{m}}|^2 
= \frac{-1}{2i}\sum_{\{1234\}}S_{\{1234\}}
\left(\bar{A}_1\bar{A}_2A_3A_4-A_1A_2\bar{A}_3\bar{A}_4\right)=0~.
\end{equation}

Combining these symmetry properties and the fact that only pairwise channels exist, there is also a conserved ``leading order energy'',
\begin{eqnarray}
& & \frac{d}{dt} \sum_n w_{n}^2 |A_{n}|^2 \\ \nonumber
&=& \frac{-1}{2i}\sum_{\{1234\}}w_{n_1}S_{\{1234\}}
\left(\bar{A}_1\bar{A}_2A_3A_4-A_1A_2\bar{A}_3\bar{A}_4\right) \\ \nonumber 
&=& \frac{-1}{4i}\sum_{\{1234\}}(w_{n_1}-w_{n_3})
S_{\{1234\}}
\left(\bar{A}_1\bar{A}_2A_3A_4-A_1A_2\bar{A}_3\bar{A}_4\right) \\ \nonumber
&=& \frac{-1}{8i}\sum_{\{1234\}}
(w_{n_1}+w_{n_2}-w_{n_3}-w_{n_4}) S_{\{1234\}}
\left(\bar{A}_1\bar{A}_2A_3A_4-A_1A_2\bar{A}_3\bar{A}_4\right)=0~.
\label{eq-numcon}
\end{eqnarray}
In the third line we used the antisymmetry when $\{12\}\leftrightarrow\{34\}$, and in the forth line we used the symmetry when $1\leftrightarrow2$ and $3\rightarrow4$. Now since the coefficient $S_{\{1234\}}$ for a resonant channel is only nonzero when the pairwise condition is met, the last line is zero. Even after including eigenstates of nonzero angular momenta, exactly the same proof goes through without change. Thus the $\phi^4$ theory, beyond spherical symmetry, also has a conserved ``leading order energy'',
\begin{equation}
\frac{d}{dt}\sum_{nl\vec{m}}w^2_{nl\vec{m}}
|A_{nl\vec{m}}|^2=0~.
\label{eq-engcon}
\end{equation} 

\subsection{Physical implications}

Note that the above two quantities are only approximately conserved. Namely, up to the leading order effect of the $\phi^4$ coupling. Among these two approximately conserved quantities, we think that the conserved particle number, Eq.~(\ref{eq-numcon}), is not very surprising. If we have a complex scalar field with $|\phi|^4$ coupling, this becomes an exactly conserved $U(1)$ current. Thus the above observation means that after limited to the real axis, the same quantity is approximately conserved. This is very similar to the fact that in flat space, if a complex scalar field theory allows exactly stable Q-balls\cite{Col85,Lee:1991ax}, then the same theory limited to the real axis allows very long-lived oscillons\cite{Bogolyubsky:1976yu,Gleiser:1993pt,CopGle95,Amin:2013ika}.

One might think that the conservation of leading order energy, Eq.~(\ref{eq-engcon}), is not surprising either. It seems to follow from the exactly conserved total energy, and the fact that energy in the coupling is further suppressed by the small field amplitude. We would like to disagree with such intuition. Remember that if there are $n$ eigenstates with nonzero amplitudes, there will only be $n$ terms in the leading order energy. However, there will generically be $n^2$ terms in the coupling energy. Thus, a conserved leading order energy means either one of the followings:
\begin{itemize}
\item Energy does not spread out too much, $n\ll|\phi^2|$, so it does not compete with the amplitude suppression.
\item Energy spreads out but there is a conspiracy in the relative phases between eigenstates such that many cross-terms vanish.
\end{itemize}
We recommend\cite{CraEvn14a,BucGre14} for further discussions about the implications to the AdS (in)stability problem.

\section{Discussion}
\label{sec-dis}

We should note that the physical origin of the ``missing top'' is a selection rule in the form of an inequality, Eq.~(\ref{eq-seln}), which is equivalent to
\begin{equation}
w_1 < \sum_{i=2}^N w_i~.
\end{equation}
This immediately reminds us that a nonlinear coupling between spherical harmonics is subjected to a similar rule about total angular momentum: the generalized triangular inequality.
\begin{equation}
l_1 \leq \sum_{i=2}^N l_i~.
\end{equation}
This suggests that there should be a group theory representation of the eigenstates such that the frequency plays the role of total angular momentum. This, together with the intriguing dependence on the parity of spatial dimensions, might be useful in proving the ``missing top'' property, or even discover further restrictions in the resonance structure, when the $\phi^N$ coupling is replaced by something more complicated.

For example, the explicit evaluation of the coupling coefficients for gravitational self-interaction, despite the simplification of spherical symmetry\cite{CraEvn14}, is already much more involved than the $\phi^4$ theory. A direct generalization beyond spherical symmetry needs to include gravitational waves\cite{HorSan14}, and it appears to be a daunting task. A group theory method would be much more preferable tool to figure out the AdS resonance structure in general.

\acknowledgments

We thank Fotios Dimitrakopoulos, Ben Freivogel, Steven Green, Luis Lehner, Matt Lippert, Javier Mas, Andrzej Rostworoski and Jorge Santos for discussions. This work is supported in part by the Foundation for Fundamental Research on Matter (FOM) of the Netherlands Organization for Scientific Research (NWO), and also the European Research Council under the European Union's Seventh Framework Programme (FP7/2007-2013) / ERC Grant agreement no.~268088-EMERGRAV.

\bibliographystyle{utcaps}
\bibliography{all_active}

\end{document}